\documentclass[10pt,twocolumn,a4paper]{article}
\usepackage{amsmath}
\usepackage{amssymb}
\usepackage{bbold}
\usepackage{graphicx}
\usepackage{caption}
\usepackage{cite}
\usepackage{color} 
\usepackage[textwidth=18.0cm,textheight=24.5cm,bottom=2cm,top=2cm]{geometry}
\usepackage{hyperref}

\begin{document}

\title{
\vspace{-1.25cm}
\textbf{\Large Theory of mechano-chemical patterning in biphasic biological tissues}
\vspace{0.0cm}
}

\author{Pierre Recho\footnote{LIPHY, CNRS UMR 5588 \&  Universit\'{e} Grenoble Alpes, F-38000 Grenoble, France.} \footnotemark[4] , Adrien Hallou\footnote{Cavendish Laboratory, Department of Physics, University of Cambridge, Cambridge, CB3 0HE, UK; Wellcome Trust/CRUK Gurdon Institute, University of Cambridge, Cambridge, CB2 1QN, UK; Wellcome Trust/MRC Stem Cell Institute, University of Cambridge, Cambridge, CB2 1QR, UK.} \footnotemark[4] and Edouard Hannezo\footnote{IST Austria, Am Campus 1, 3400 Klosterneuburg, Austria.} \footnote{A.H, P.R and E.H contributed equally to this work. E-mail: pierre.recho@univ-grenoble-alpes.fr, ah691@cam.ac.uk or edouard.hannezo@ist.ac.at}}

\date{}

\maketitle

\renewcommand{\abstractname}{\vspace{-\baselineskip}}
\begin{abstract}
\textbf{\small The formation of self-organized patterns is key to the morphogenesis of multicellular organisms, although a comprehensive theory of biological pattern formation is still lacking. Here, we propose a minimal model combining tissue mechanics to morphogen turnover and transport in order to explore new routes to patterning. Our active description couples morphogen reaction-diffusion, which impact on cell differentiation and tissue mechanics, to a two-phase poroelastic rheology, where one tissue phase consists of a poroelastic cell network and the other of a permeating extracellular fluid, which provides a feedback by actively transporting morphogens. While this model encompasses previous theories approximating tissues to inert monophasic media, such as Turing's reaction-diffusion model, it overcomes some of their key limitations permitting pattern formation via any two-species biochemical kinetics thanks to mechanically induced cross-diffusion flows. Moreover, we describe a qualitatively different advection-driven Keller-Segel instability which allows for the formation of patterns with a single morphogen, and whose fundamental mode pattern robustly scales with tissue size. We discuss the potential relevance of these findings for tissue morphogenesis.}
\end{abstract}

\maketitle

\vspace{0.2cm}

How symmetry is broken in the early embryo to give rise to a complex organism, is a central question in developmental biology. To address this question, Alan Turing proposed an elegant mathematical model where two reactants can spontaneously form periodic spatial patterns through an instability driven by their difference in diffusivity \cite{Turing1952}.
Molecular evidence of such a reaction-diffusion scheme \textit{in vivo} remained long elusive, until pairs of activator-inhibitor morphogens were proposed to be responsible of pattern formation in various embryonic tissues \cite{Gierer1972,Murray2003,Kondo2010, Sick2006,Economou2012,Inomata2013,Muller2012,Raspopovic2014}. 
Interestingly, these studies also highlight some theoretical and practical limitations of existing reaction-diffusion models, including the fact that Turing patterns require the inhibitor to diffuse at least one order of magnitude faster than the activator ($D_{I}/D_{A}> 10$) \cite{Murray2003}, although most morphogens are small proteins of similar molecular weights, implying that $D_{I}/D_{A} \approx 1$. As a consequence, the formation of Turing patterns \textit{in vivo} should result from other properties of the system such as selective morphogen immobilisation \cite{Castets1990,Lengyel1991,Rauch2004} or active transport \cite{Rovinsky1993} as demonstrated in synthetic sytems. Moreover, reaction-diffusion models of pattern formation entail a number of restrictions regarding the number and interactions of morphogens, and pattern scaling with respect to the tissue size, which have been all limiting their quantitative applicability \textit{in vivo}. While the genetic and biochemical aspects of developmental pattern formation have been the focus of most investigations, the interplay between mechanics and biochemical processes in morphogenesis started to unfold following some pioneering contributions \cite{Oster1983}. The crucial role played by multiphasic tissue organisation and active cell behaviours in biological pattern formation is now an active field of research  \cite{Bois2011,Howard2011, Weber2011, Hiscock2015}.

In this article, we derive a general mathematical formulation of tissues as active biphasic media coupled with reaction-diffusion processes, where morphogen turnover inside cells, import/export at the cell membrane and active mechanical transport in the extracellular fluid are coupled together through tissue mechanics. While encompassing classical reaction-diffusion results \cite{Turing1952,Gierer1972,Murray2003,Kondo2010}, for instance allowing import-export mechanisms to rescale diffusion coefficients and to form patterns with equally diffusing morphogens \cite{Rauch2004}, this theory provides multiple new routes to robust pattern formation. In particular, assuming a generic coupling between intracellular morphogen concentration and poroelastic tissue mechanics, we demonstrate the existence of two fundamentally different non-Turing patterning instabilities, respectively assisted and driven by advective extracellular fluid flows, explaining pattern formation with only a single morphogen with robust scaling properties, and how patterning can be independent of underlying morphogen reaction schemes. Finally, we discuss the biological relevance of such a model, and in particular its detailed predictions that could be verified \textit{in vivo}. 

\vspace{-0.2cm}

\subsection*{Derivation of the model}

\begin{figure*}[!htp]
\begin{center}
\includegraphics[width=17.5cm]{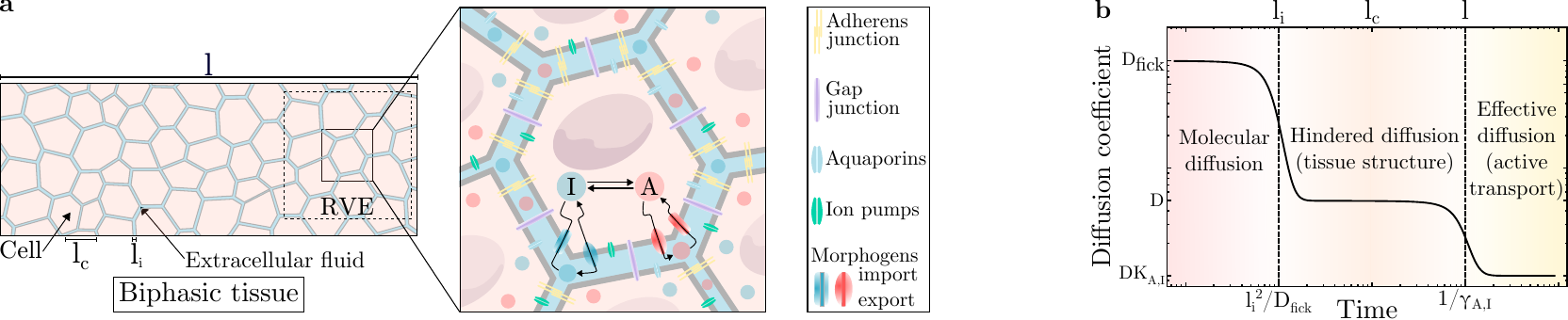}
\end{center}
\vspace{-10pt}
\caption{\textbf{Model for pattern formation in active biphasic tissues.} (\textbf{a}) Schematic of the model: (Left) Cells form a poroelastic network, permeated by extracellular fluid, where three natural length scales can be defined: the interstitial space size ($l_i$), the characteristic cell size ($l_c$) and the tissue size ($l$). (Right)  Biochemical interactions between morphogens, $A$ and $I$, take place inside the cell and are described by their respective turnover rate functions $f(A,I)$ and $g(A,I)$. $A$ and $I$ are exported across the cell membrane at rates $\lambda_{A,I}$ and imported at rates $\gamma_{A,I}$, respectively. In the extracellular space, both $A$ and $I$ spread freely by diffusion at the same rate $D$, or can be advected by the fluid at velocity $v_e$. (\textbf{b}) Evolution of the effective diffusion coefficient as a function of time and space scales. At shorter distances and times, diffusive behaviour of morphogens is described by a molecular diffusion coefficient, $D_{\text{Fick}}$. At intermediate scales, the diffusive motion of morphogens starts to be hindered by cells and the global diffusion coefficient, $D$, depends of the tissue spatial organisation through $\phi^*$. At larger scales, morphogen diffusion is controlled by dynamic interactions with cells (import/export, adsorption/desorption,) and an effective coefficient $DK_{A,I}$ \cite{Muller2012}.}
\vspace{-5pt}
\label{Fig1}
\end{figure*}

As sketched in Fig.~\ref{Fig1}(a), we model multicellular tissues as continuum biphasic porous media of typical length $l$, with a first phase consisting of a poroelastic network made of adhesive cells of arbitrary shape and typical size $l_c$ (with local volume fraction $\phi$), and a second phase of aqueous extracellular fluid permeating in-between cells in gaps of a characteristic size $l_i$. These two internal length scales disappear in the coarse-graning averaging over a representative volume element of typical lengthscale $l_r$ satisfying $l_{i,c}\ll l_r\ll l$. Both phases are separated by cell membranes, actively regulating the interfacial exchange of water and other molecules thanks to genetically controlled transport mechanisms \cite{Bokel2014,Kicheva2012}. At the boundary of the domain, no-flux boundary conditions are imposed such that the system is considered in isolation. We present below the main steps of the model derivation, which are detailed in SI Appendix.

\subsubsection*{Intracellular morphogen dynamics}
 
Morphogens enable cell-cell communication across the tissue and determine cell fate decisions. Importantly, most known morphogens cannot directly react together and as such, have to interact ``through'' cells (or cell membranes) where they are produced and degraded \cite{Kicheva2012}.
Concentration fields of two morphogens, $A_{i,e}(\vec{r},t)$ and $I_{i,e}(\vec{r},t)$, are thus defined separately in each phase of the system, indices $(i,e)$ denoting intra- and extra-cellular phases, respectively. The conservation laws of the intracellular phase, which cannot be transported, read:
\begin{equation}\label{e:chem_cells}
\begin{array}{c}
\partial_t(\phi {A}_i)= f({A}_i,{I}_i)+\gamma_A  {A}_e-\lambda_A {A}_i\\
\partial_t(\phi{I}_i)= g({A}_i,{I}_i)+\gamma_I {I}_e-\lambda_I {I}_i
\end{array}
\end{equation}
where $\partial_t$ denotes the partial derivative with respect to time and $\gamma_{A,I}$ (resp. $\lambda_{A,I}$) the import (resp. export) rates of morphogens (which can also describe immobilization rates at the cell membrane). We also introduce $f$ and $g$, the non-linear morphogen turnover rates describing their production and degradation by cells, with a single stable equilibrium solution $f(A_i^*,I_i^*)=g(A_i^*,I_i^*)=0$. Finally, we introduce the transmembrane transport equilibrium constants by $K_A=\lambda_A /\gamma_A$ and $K_I=\lambda_I / \gamma_I$. Although the import/export coefficients $K_{A,I}$ could in principle depend on morphogen concentrations, this constitutes a non-linear effect that we ignore in our linear theory.

\subsubsection*{Extracellular fluid dynamics}

Next, we write a mass conservation equation for the incompressible fluid contained in the tissue interstitial space between cells: 
\begin{equation}\label{e:water_conservation}
\begin{array}{c}
\partial_t\phi-\nabla.((1-\phi)v_e)  = \frac{\phi_{h}({A}_i,{I}_i)-\phi}{\tau}
\end{array}
\end{equation}
where $v_e$ is the velocity of the extracellular fluid.  The right-hand side of this equation describes the fact that cells actively regulate their relative volume fraction to an homeostatic value $\phi_{h}({A}_i,{I}_i)$ at a timescale $\tau$ \cite{Hoffmann2009}. Note that, \eqref{e:water_conservation} with $v_e \neq 0$ implies a recirculation of internal fluid,  via gap junctions \cite{Zehnder2015} (SI Appendix, Sec.~1.A.3).

As detailed below, we assume that local cellular morphogen concentrations have an influence on the volume fraction $\phi$ which couples tissue mechanics to local morphogens concentration in our theory.
At linear order, this coupling generically reads $\phi_h({A}_i,{I}_i)= \phi^*+\chi_A ({A}_i-{A}_i^*)/{A}_i^*+\chi_I({I}_i-{I}_i^*)/{I}_i^*$ where we denote $\phi^*=\phi_h({A}_i^*,{I}_i^*)$, the equilibrium cell volume fraction, and the $\chi_{A,I}$ terms account for the sensitivity of cell volume to intracellular morphogen concentrations. Such a mechano-chemical effect on the tissue packing fraction, $\phi$, can occur either via the active control of individual cell volume \cite{Hoffmann2009} or through the active balance between cell proliferation and loss (SI Appendix, Sec.~1.A.4), with $\chi_{A,I}>0$ for morphogens acting as growth factors and $\chi_{A,I}<0$ for morphogens working as growth inhibitors. 
This is a reasonable assumption, as a number of morphogens involved in cell fate decisions can act as growth factor/inhibitors \cite{Smith1981,Ginzberg2015}, and \textit{in vitro} experiments have shown that cells, upon exposure to factors such as FGF or EGF, elicit a series of signaling mediated responses involving an increase in transmembrane ion flux, cell volume changes \cite{Hoffmann2009} and subsequent cell growth/division \cite{Zetterberg1984}.
Moreover, during digits pattern formation in the limb bud, which has been proposed to rely on a Turing instability, morphogens such as BMP participate in both the reaction-diffusion scheme \cite{Raspopovic2014} and in morphogenetic events such as cell condensation \cite{Benazet2012}, with skeletal formation being associated with large cell volume fraction changes \cite{Cooper2013}. The cell volume fraction is thus highly modulated in space and time, concomitantly with morphogen pattern formation \cite{Benazet2012}, advocating for the need of a global mechano-chemical theory taking into account both effects.

\subsubsection*{Extracellular morphogen dynamics}

Morphogens, once secreted by cells, are transported by diffusion and advection in the extracellular fluid:
\begin{equation}\label{e:transport_ext_av}
\small
\begin{array}{c}
\partial_t((1-\phi){A}_e)+\nabla.\left( (1-\phi){A}_ev_e-D\nabla {A}_e \right) =-\gamma_A  {A}_e+\lambda_A {A}_i\\
\partial_t((1-\phi){I}_e)+\nabla.\left((1-\phi){I}_e v_e-D\nabla {I}_e \right)=-\gamma_I  {I}_e+\lambda_I {I}_i
\end{array}
\end{equation}
where $D$ is the global Fickian diffusion coefficient of both morphogens depending on tissue packing and tortuosity \cite{Crank1979, Bear1989, Muller2012}. As we are interested in a linear theory, we consider here $D=D(\phi^*)$ as a constant. We neglect here, for the sake of  simplicity, phenomena such as extracellular morphogen degradation or the influence of  extracellular morphogen concentrations on reaction terms, as they do not modify qualitatively the dynamics (SI Appendix, Sec.~1.C). Note that one could also take into account, at the mesoscopic level, some effective non-local interactions such as cell-cell communication via long-ranged cellular protrusions \cite{Kondo2017}. This may require to consider spatial terms in \eqref{e:chem_cells} to introduce an additional characteristic lengthscale from non-local cell-cell transport.

\subsubsection*{Mechanical behaviour of the cellular phase}

To complete our description, we need to specify a relation linking cell volume fraction to interstitial fluid velocity. For this, we use a poroelastic framework, whose applicability to describe the mechanical response of biological tissues has been thoroughly investigated in various contexts \cite{Netti2000,Fraldi2018}. Taking an homogeneous tissue as reference state, poroelastic properties imply that a local change of the cell volume fraction creates elastic stresses in the cellular phase which translate to gradients of extracellular fluid pressure $p$. Such gradients of pressure in turn drive extracellular fluid flows, which can advect morphogens, and we show (SI Appendix, Sec.~1.A.7) that this effects results in a simple Darcy's law between cell volume fraction and fluid flow \cite{Bear1989}:
\begin{equation}\label{e:pressure_vol_frac}
\begin{array}{c}
(1-\phi)v_e=-\frac{\kappa}{\eta}\nabla p= D_m \nabla \phi.
\end{array}
\end{equation}
This relation introduces the hydrodynamic diffusion coefficient of the extracellular fluid, $D_{m}=K\kappa/\eta$, a key mechanical parameter of the model which feeds back on the reaction diffusion dynamics \eqref{e:transport_ext_av}, with $\kappa$ the tissue permeability, $K$ the elastic drained bulk modulus and $\eta$ the fluid viscosity. The hydrodynamic length scale $l_m=\sqrt{D_m \tau}$ is associated to such fluid movement. Importantly, we only explore here the simplest tissue rheology for the sake of simplicity and concision. Nevertheless, we also investigate (SI Appendix, Sec.~1.H) the role of growth and plastic cell rearrangements and show that they can be readily incorporated in our model, leading to different types of patterning instabilities. However, we would like to highlight here that the results presented thereafter are all robust to small to intermediate levels of tissue rearrangements.

\subsection*{Model of an active biphasic tissue}

Eqs.(\ref{e:chem_cells}-\ref{e:pressure_vol_frac}) define a full set of equations describing the chemo-mechanical behaviour of an active biphasic multicellular tissue (SI Appendix, Sec.~1.B). To provide clear insights on the biophysical behaviour of the system, we focus on a limit case where $\gamma_{A,I}\gg \lambda_{A,I} \gg f,g$ such that $K_{A,I}\ll 1$. This corresponds to an ubiquitous biological situation where rates of membrane transport are order of magnitudes faster than transcriptionaly controled morphogen turnover rates, and where endocytosis occurs at a much faster rate than exocytosis. In that case, the relations ${A}_e\simeq K_A A_i$ and ${I}_e\simeq K_I I_i$ always hold and even if a significant fraction of morphogens is immobilized inside the cells \cite{Muller2012}, the import/export terms cannot be neglected as $\gamma_{A,I}$ are very large, so that $\gamma_A(A_e-K_AA_i)$ and $\gamma_I(I_e-K_II_i)$ are indeterminate quantities (SI Appendix, Sec.~1.C). Summing both internal \eqref{e:chem_cells} and external \eqref{e:transport_ext_av} conservation laws, we obtain a simplified description of the system (SI Appendix, Sec.~1.C): 
\begin{equation}\label{e:simplified_model}
\begin{array}{c}
\partial_t(\phi {A}_i)+\nabla.\left(A_iK_AD_m\nabla \phi-K_AD\nabla A_i \right)=f(A_i,I_i)\\
\partial_t(\phi {I}_i)+\nabla.\left(I_i K_ID_m\nabla \phi-K_ID\nabla I_i\right)=g(A_i,I_i)\\
-l_m^2 \Delta \phi +\phi  =\phi_h({A}_i,{I}_i).
\end{array}
\end{equation} 

Non-dimensionalizing times with $\tau_A$ associated with the degradation of $A_i$ in the morphogen turnover functions $f$ and $g$ and lengths with $l_A=\sqrt{K_A D \tau_A}$ we find that \eqref{e:simplified_model} is controlled by a few non-dimensional parameters: $K_I/K_A$ describes the mismatch of morphogen membrane transport, $D_m/D$ compares the global hydrodynamic and Fickian diffusion of the morphogens, $\tau/(K_A \tau_A)$ compares the response time of cell volume fraction to the effective morphogen turnover rate, and $\chi_A$ and $\chi_I$ account for the sensitivity of $\phi$ to morphogen levels. Using this restricted set of parameters encapsulating the behaviour of the model, we investigate several of its biologically relevant limits, demonstrating that they provide independent routes towards tissue patterning. 

\subsection*{Orders of magnitude on morphogen transport}

In the simplest limit of the model, the cell fraction remains constant, $\phi=\phi^*$, which is valid if the effect of the morphogens on $\phi$ is very small compared to the restoring mechanical forces (i.e. $\chi_{A,I}= 0$). The model then reduces to Turing's original system, with diffusion coefficients being renormalised by morphogens transmembrane transport equilibrium constants, $K_{A,I}D$, similar to results obtained in \cite{Rauch2004, Muller2012}. This implies that even species with similar $D$, can exhibit effective diffusion coefficients widely differing from each other on longer timescales and produce Turing patterns when $K_I\gg K_A$ (SI Appendix, Sec.~1.F).

\begin{figure*}[!ht]
\begin{center}
\includegraphics[width=15cm]{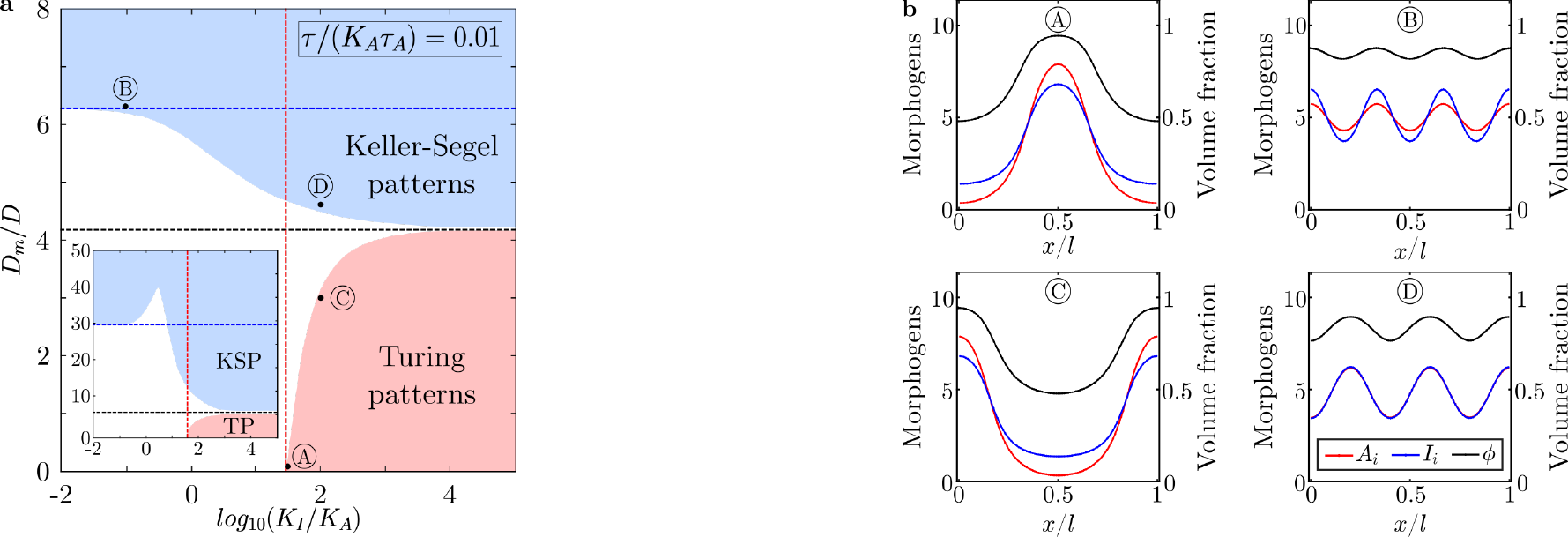}
\end{center}
\vspace{-10pt}
\caption{\textbf{Linear stability analysis and numerical simulations of pattern formation in active biphasic tissues.} (\textbf{a}) Phase diagram of \eqref{e:simplified_model} in the $(K_I/K_A, D_m/D)$ parameter space for $\tau/(K_A\tau_A)=0.01$ and $\tau/(K_A\tau_A)=0.1$ (inset). The red and blue dashed lines correspond to analytical thresholds of instability (given in the text) for Turing  and Keller-Segel patterns respectively. The black dashed line is the analytical phase boundary between both regimes in the limit $K_I\gg K_A$ given by $\chi_A =D/D_m+\tau/(\tau_AK_A)$. This limit is shifted up when the ratio $\tau / \tau_A K_A$ is increased, while a pronounced notch appears in the ``Keller-Segel patterns'' domain (see inset). Other parameters are set to $\chi_A=0.25$, $\chi_I=0$,  $\tau_I/(K_A\tau_A)=0.2$, $K_A\tau_A \rho=1$, $\phi^*=0.85$ and large tissue size ($l_A/l\ll 1$). (\textbf{b}) 1D numerical simulations of \eqref{e:simplified_model} with random initial conditions for several choices of parameters identified by letters A, B, C \& D.  $l_A/l=0.1$. }
\vspace{-10pt}
\label{Fig2}
\end{figure*}
 
In Fig.~\ref{Fig1}(b), we depict scaling arguments for the changes in effective diffusion coefficient at various time/length scales, associated both with tissue structure and import/export kinetics \cite{Rauch2004}.
At small timescales, diffusion is characterised by a local Fickian diffusion coefficient, theoretically expected to be of the order of $D_{\text{Fick}}\approx 10^{-11} \text{m}^2 \text{s}^{-1}$, in line with fluorescence correlation spectroscopy (FCS) measurements \cite{Kicheva2012,Inomata2013,Muller2012}.
This occurs across a typical cell-to-cell distance of $l_i \approx 10^{-7}-10^{-9}$m \cite{Barua2017}, so that this regime is valid for time scales below $l_i^2/D_{\text{Fick}} \approx 10^{-2}-10^{-6}$s, which is much faster than the typical import/export kinetics of $1/ \gamma_{A,I} \approx 10^1-10^2$s \cite{Smith1996}. 
At intermediate timescales, the diffusion coefficient needs to be corrected for volume exclusion effects due to the porous nature of the tissue, an effect which can be very large for cell volume fraction close to one \cite{Blassle2018}. An upper bound (Hashin-Shtrikman) for global diffusion can be computed, irrespective of the microscopic details of tissue geometry, as $D(\phi^*)\leq D_{\text{Fick}}(1-\phi^*)/(1+\phi^*/2)$ \cite{Crank1979}, which would suggest, in the case of $\phi^* \approx 0.8-0.9$, that it should be around an order of magnitude smaller than local diffusion, $D (\phi^*) \approx 10^{-12} \text{m}^2 \text{s}^{-1}$.
Finally, at the time scales larger than $1/ \gamma_{A,I}$ described by the present model, the diffusion is decreased further by a factor $K_{A,I}$, i.e. by the relative concentrations of morphogens ``trapped" cellularly (i.e. a $1-10$ ratio) such that $D(\phi^*) K_{A,I}\approx 10^{-12}-10^{-13}\text{ m}^2\text{s}^{-1}$. 
This is consistent with effective diffusion coefficients  measured from tissue-wide fluorescence recovery after photobleaching (FRAP) over minutes to hours time scales \cite{Kicheva2012,Muller2012,Inomata2013,Blassle2018}. Note here, that the respective contributions of volume exclusion and import/export effects on FRAP measured diffusion coefficients are non-trivial and are detailed in SI Appendix, Sec.~1.H.
Overall, although our model in its simplest limit ($\phi=\phi^*$) relaxes the classical Turing condition $D_{I} \gg D_{A}$, it still implies quite stringent conditions on the ratio of intracellular and extracellular morphogens ($I_{e}/I_{i} \gg A_{e}/A_{i}$). Exploring further the effect of a variable cell volume fraction $\phi$, we demonstrate that coupling morphogen dynamics and tissue mechanics suppresses this limitation via active transport of morphogens.

\subsection*{Turing-Keller-Segel instabilities}

To assess the regions in parameter space where stable patterns can form in our mechano-chemical framework, we perform a linear stability analysis on \eqref{e:simplified_model}. Here, we consider a classical Gierer-Meinhardt activator-inhibitor scheme \cite{Gierer1972}: $f(A,I)=  \rho A^2/I - A/\tau_A$ and $g(A,I)= \rho A^2 - I/\tau_I$, where $\rho$ is the rate of activation and inhibition and $\tau_{A,I}$ the timescales of degradation of $A$ and $I$ \cite{Gierer1972} and the particular case of a single morphogen capable of increasing $\phi_h$ ($\chi_A>0,\chi_I=0$).

\begin{figure}[t]
\begin{center}
\includegraphics[width=8.0cm]{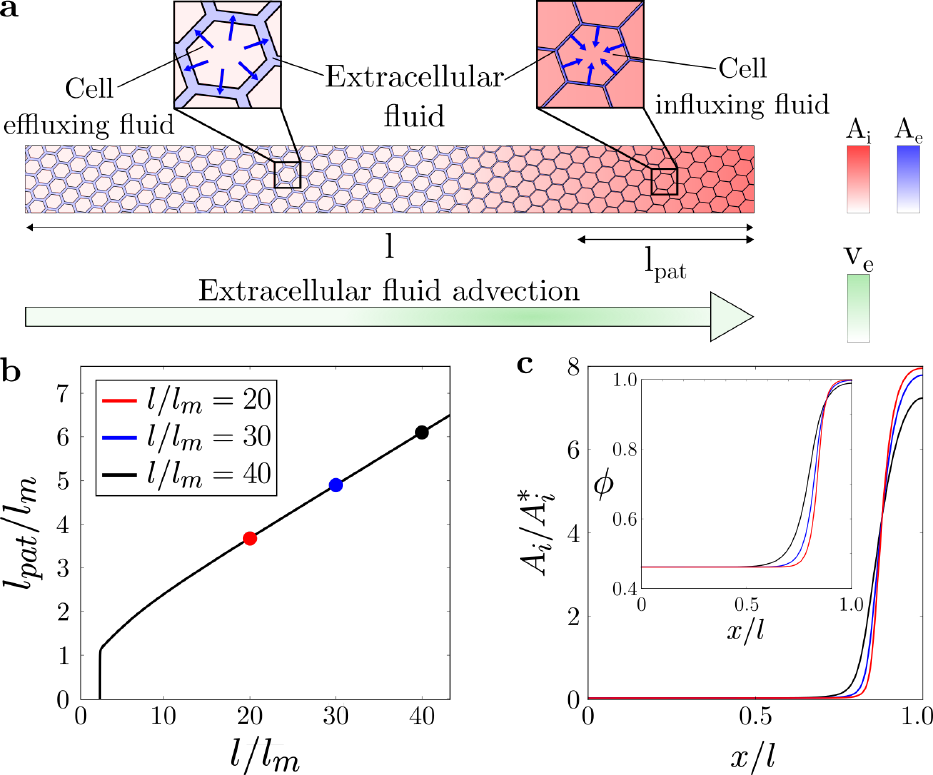}
\end{center}
\vspace{-10pt}
\caption{\textbf{Scaling properties of the Keller-Segel instability with one morphogen.}(\textbf{a}) Schematic of the Keller-Segel instability in a 1D tissue. Morphogens gradients generate cell volume fraction gradients (via local fluid exchanges, blue arrows in inset), which in return cause mechanically-induced self-amplifying extracellular flows that advect morphogens from morphogen-poor to morphogen-rich regions (green arrow). (\textbf{b}) Normalized pattern size as a function of system size in the single morphogen case with $f=0$. (\textbf{c}) Morphogen concentration and cell packing fraction (inset) profiles remain quasi-stationary as system size increases. Parameters are $\chi_A=0.25$, $D_m/D=10$ and $\phi^*=0.85$ }
\vspace{-10pt}
\label{Fig3}
\end{figure}

In the phase diagram in Fig.~\ref{Fig2}~(a), we show that two distincts instabilities can be captured by this simplified theory. 
The first instability, identified here as ``Turing patterns'', corresponds to a classical Turing instability, where diffusive transport of morphogens dominates over their advection by interstitial fluid ($D_m\ll D$) and with instability threshold given by $K_I\tau_I-K_A\tau_A>2\sqrt{\tau_A\tau_IK_AK_I}$ for $l_A/l\ll 1$(dashed red line on Fig.~\ref{Fig2}~(a)) which, as expected, is always true regardless of the value of $\tau_{A,I}$ if $K_I\gg K_A$. 
However, another generic pattern forming instability driven by active transport phenomena  is present in the phase diagram, labelled ``Keller-Segel patterns'' \cite{Keller1970}.
The physical origin of the resulting pattern is here similar to active fluid instabilities \cite{Bois2011,Recho2013,Recho2015,Hannezo2015,Weber2011,Aguilar2018}: if stochastic local changes in morphogen concentration result in an increase in cell volume fraction, fluid must be pumped inside cells. This causes local elastic deformations in the tissue which generate large-scale extracellular fluid flows from regions of low to high morphogen concentration, resulting in a positive feedback loop of morphogens enrichement (Fig.~\ref{Fig3}~(a)), and steady-state patterns. Interestingly, such an instability can even occur for a single morphogen.
In this limit, patterning occurs if $\sqrt{\chi_A} > \sqrt{D/D_m}+\sqrt{ \tau/(\tau_AK_A)}$ when $l_A/l\ll 1$ so that the volume fraction sensitivity $\chi_A$  is above a critical value (dashed blue line in Fig.~\ref{Fig2}~(a), which captures well the phase boundary in the limit $K_A\gg K_I$, although the instability occurs generically for any value of $K_{A,I}$).
The number of patterns displayed by the profiles shown on Fig.~\ref{Fig2} (b) can be predicted by linear analysis (See Appendix, Sec.~1.D) because they are chosen close to the onset of instability.

Thus, coupling tissue mechanical behaviour to morphogen reaction-diffusion provides, via the generation of advective fluid flows, a new route to stable pattern formation with a single morphogen. Moreover, this instability has two remarkable features. First, it only requires the presence of a single morphogen (SI Appendix, Sec.~1.G) which could correspond to many practical situations where a pair of activator/inhibitor has not been clearly identified, for instance the role of Wnt in the antero-posterior pattern of planarians \cite{Stuckemann2017}. Second, it possesses spatial scaling properties regarding to its fundamental mode, as compared to a Turing instability. Indeed, when morphogen turnover rate is small compared to its effective hydrodynamic and Fickian diffusion ($f \to 0$), the fundamental mode, i.e. a single two-zones pattern, is the most unstable in a robust manner, given that morphogen turnover $f$ stabilises specifically this mode (SI Appendix, Sec.~1.G.2), whereas in the case of a Turing instability, this would require fine-tuning and marginally stable reaction kinetics. We illustrate such a scaling property in Fig.~\ref{Fig3}.
This mechanism could potentially apply to situations where a binary spatial pattern is independent of system size such as dorso-ventral or left-right patterns in early vertebrate embryos \cite{Inomata2013,Muller2012}, or planarian antero-posterior pattern \cite{Stuckemann2017,Werner2015}. If so, it could provide a simpler alternative to previously proposed mechanisms involving additional species or complex biochemical signaling pathways \cite{Inomata2013,Werner2015}. 

Importantly, simple estimates can be used to demonstrate the biological plausibility of such mechanical effects during morphogenetic patterning. A key parameter driving Keller-Segel instabilities is the hydrodynamic diffusion coefficient $D_m$, which can be estimated from values of the drained bulk modulus $K\approx 10^4$ Pa \cite{Netti2000} and the tissue permeability upper bound \cite{Crank1979} $\kappa\approx l_i^2(1-\phi^*)/(1+\phi^*/2)$ with $l_i\approx 10^{-7}-10^{-9}$m and $\phi^*\approx 0.85$ as above. Using $\eta\approx 10^{-3}$ Pa.s (water viscosity), we obtain $D_m\approx 10^{-12}-10^{-8}\text{ m}^2\text{s}^{-1}$, showing that the  hydrodynamic diffusion can be similar or even much larger than Fickian diffusion. In agreement with typical timescales involved in regulatory volume increase or decrease of cells following an osmotic perturbation \cite{Hoffmann2009}, we estimate that $\tau \approx 10^2$ s, while morphogen turnover time scale has been measured as $\tau_A\approx 10^4-10^5$ s \cite{Muller2012}. With $K_A\approx 0.1$ as above, we obtain $\tau/(K_A\tau_A)\approx 0.01-0.1$, which is used in Fig.~\ref{Fig2}, and displays broad regions of instability, although parameters like sensitivities $\chi_{A,I}$ would need to be better assessed \textit{in vivo} in future works.

\subsection*{Cross-diffusion Turing instabilities}

Finally, we investigate the behaviour of our model (\eqref{e:simplified_model}), when cell fraction sensitivity to morphogen concentration is negative ($\chi_{A,I}<0$), eliminating the possibility of up-hill morphogen diffusion at the origin of the Keller-Segel instability. We also consider that $f$ and $g$ do not necessarily follow an activator-inhibitor kinetics, but any possible interaction scheme between two morphogens. 
For mathematical clarity on the physical nature of the instability studied here, we make the simplifying assumptions that $\tau=0$ and $\chi_{A,I} \ll 1$, with $D \sim D_{m} \chi_{A,I}$ in \eqref{e:simplified_model}. This relates to a realistic biological situation, where cell volume fraction relaxes rapidly after perturbation and depends weakly on morphogen levels, yielding:
\begin{equation}\label{e:cross_diff_simple}
\begin{array}{c}
\phi^*\partial_t{A}_i+\nabla.\left(A_iK_AD_m\nabla \phi_h-K_AD\nabla A_i \right)=f(A_i,I_i)\\
\phi^*\partial_t{I}_i+\nabla.\left(I_i K_ID_m\nabla \phi_h-K_ID\nabla I_i\right)=g(A_i,I_i).
\end{array}
\end{equation} 

In this limit, the conditions for linear stability of the homogeneous solution are exactly the ones of a  classical Turing system but with cross-diffusion terms (SI Appendix, Sec.~1.E). Such a scenario has been studied in the framework of monophasic reaction-diffusion systems with \textit{ad hoc} cross-diffusion terms \cite{Madzvamuse2015}, which arise generically in various chemical and biological systems \cite{Vanag2009}. Our work thus provides a particular biophysical interpretation of these terms in multicellular tissues, which we show to originate from intrinsically mechano-chemical feedbacks between morphogen dynamics and tissue mechanics.

\begin{figure}[!hb]
\begin{center}
\includegraphics[width=8.5cm]{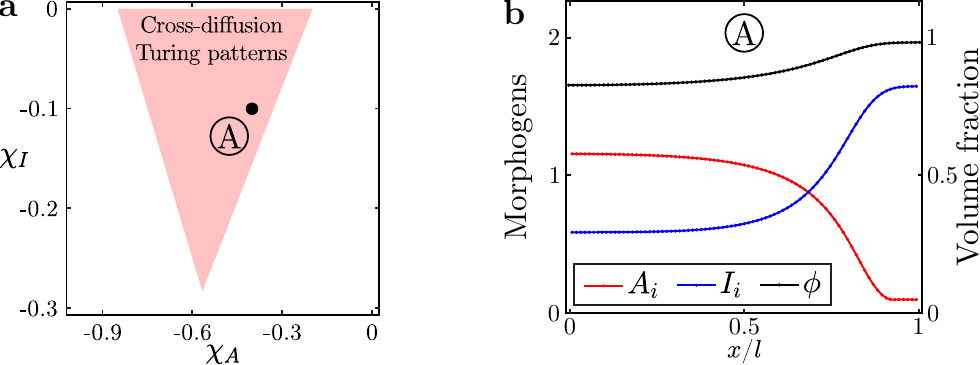}
\end{center}
\vspace{-10pt}
\caption{\textbf{Pattern formation for cross-diffusion Turing instabilities.} (\textbf{a}) Phase diagram of \eqref{e:simplified_model} in the $(\chi_A, \chi_I)$ space obtained by numerical linear stability analysis. Parameters are $\tau/(K_A\tau_A)=0.01$, $D_m/D=10$, $K_I/K_A=10$, $\tau_I/(K_A\tau_A)=0.9$, $\phi^*=0.85$ and $l_A/l\ll 1$. (\textbf{b}) 1D numerical simulation of \eqref{e:simplified_model} using a simple inhibitor-inhibitor reaction scheme (SI Appendix, Sec. 1.B). }
\vspace{-10pt}
\label{Fig4}
\end{figure}

As shown in \cite{Madzvamuse2015}, such cross diffusion terms result in  a dramatic broadening of the phase space for patterns. In particular, any two-morphogen reaction scheme can now generate spatial patterns and not just the classical activator-inhibitor schemes. For instance, it becomes possible to obtain patterns with activator-activator or inhibitor-inhibitor kinetics similar to those observed in numerous gene regulatory networks or signaling pathways involved in cell fate decisions \cite{Zhou2011}. 
We illustrate this result by considering an inhibitor-inhibitor kinetic scheme, which cannot yield patterns in the classical Turing framework, wandemonstrate analytically and numerically the existence of a  region of stable patterns (from \eqref{e:simplified_model}), where a cross-diffusion driven Turing instability can develop (Fig.~\ref{Fig4}).

\section*{Discussion}
  
In this paper, we have introduced a generalisation of Turing's work on pattern formation in biological tissues by coupling equations describing the structure and mechanical properties of multicellular tissues with a classical reaction-diffusion scheme. In particular, our work highlights two important features of multicellular tissues, as of yet largely unexplored in this context: their biphasic nature, i.e. the fact that morphogen production/degradation is controlled by  cells while transport takes place extracellularly requiring active membrane exchanges (effectively rescaling diffusion \cite{Rauch2004,Muller2012}), and the possibility for active large scale flows to develop within the tissue interstitial space. We demonstrate that coupling tissue cell volume fraction to local morphogen levels (based on the dual role of morphogens in patterning and cell growth/volume regulation \cite{Smith1981,Ginzberg2015}) provides a biophysically realistic route towards two qualitatively different modes of patterning instability. Extracellular fluid flows can have two important consequences on patterning. Firstly, as the Turing instability is rooted in the cross-effects  between a stable chemical reaction of two morphogens and their diffusion, the conditions of such instability are deeply affected by active hydrodynamic transport which can create cross terms into the effective diffusion matrix. This causes a drastic widening of the phase space of Turing patterning, rendering it robust and only weakly dependent on morphogens reaction scheme. Secondly, extracellular fluid flows can also create an instability of a different nature (Keller-Segel), when these flows have an anti-diffusive structure, spontaneously creating morphogens gradients. Here, chemical reactions between morphogens are only setting the number of patterns, and if such reactions are sufficiently slow, the spatial pattern of morphogen always coarsens to the fundamental mode of instability, and has robust scaling properties compared to conventional Turing models. This could have interesting implications concerning recent experimental evidences for robust scaling of the Nodal/Lefty pattern in the early zebrafish embryo \cite{Almuedo2018}.

In this respect, our approach, which has the advantage of parsimony, taking into account the manifest biphasic nature of multicellular tissues, is complementary to others which have been proposed to solve limitations of Turing's model by introducing additional morphogen regulators \cite{Werner2015,Diego2018}, and also displays connections with recent development in the mechano-chemical descriptions of active fluids such as the cell cytoskeleton \cite{Bois2011,Howard2011}. Nevertheless, although our hypothesis of cell volume fraction gradients driving large-scale flows is generic to biphasic tissues,  further quantitative experiments would be needed to test the relationship between morphogen concentration and cell volume fraction, as well as probe the role of transmembrane import/export kinetics or similar phenomena such as transmembrane signaling \cite{Rauch2004}, morphogen adsorption/desorption on cell surface \cite{Muller2012} and long-distance cellular protrusions \cite{Kondo2017}, on effective morphogen diffusion rates. Systems such as digits patterning, where cell volume fraction spatial pattern appears concomitant to morphogen patterns \cite{Benazet2012}, or planarian antero-posterior patterning, where pairs of activator/inhibitor have not been clearly identified \cite{Stuckemann2017}, provide possible testing grounds for our model. Interestingly, large-scale extracellular fluid flows have been increasingly observed during embryo development, not only in the classical case of cilia driven flows \cite{Freund2012}, but also due to mechanical forces arising from cellular contractions as well as osmotic and poro-viscous effects \cite{Krens2017, Ruiz2017}, calling for a more systematic understanding of passive vs. active transport mechanisms during embryonic pattern formation. Whether biological examples of Turing patterning instabilities, such as left-right or dorso-ventral patterning, digits pattern formation or skin appendages patterns are causally associated with concomitant changes in cell volume and/or cell packing remains a result to be experimentally investigated.

\section*{Methods}
Linear stability analysis was performed numerically using Mathematica, while numerical integrations of the model equations were performed using a custom-made Matlab code.

\section*{Acknowledgments}
A.H acknowledges the University of Cambridge for a David Crighton Fellowship and the Wellcome Trust for an Interdisciplinary Research Fellowship. P.R. acknowledges a CNRS-Momentum grant and E.H from the Austrian Science Fund (FWF) [P 31639]. The authors warmly thank Benjamin Simons, Anna Kicheva, David J\"org, Tom Hiscock, Pau Formosa-Jordan, Erik Clark and Lev Truskinovsky for useful discussions.


\end{document}